\documentclass[10pt]{article}
\usepackage[OE]{express}
\usepackage{siunitx}

\begin{document}

\title{Coherent perfect absorption in a weakly absorbing fiber}

\author{Ali. K. Jahromi and Ayman F. Abouraddy\authormark{*}}

\address{CREOL, The College of Optics \& Photonics, University of Central Florida, Orlando, Florida 32816, USA}

\email{\authormark{*}raddy@creol.ucf.edu} 



\begin{abstract}
Coherent perfect absorption refers to the interferometric enhancement of absorption in a partially lossy medium up to 100\%. This can be achieved without modifying the absorbing medium itself by instead engineering its photonic environment. Ion-doped fibers are one of the most technologically relevant absorbing materials in optics, which are widely employed in fiber amplifiers and lasers. Realizing complete optical absorption of an incident field in short-length moderately-doped fibers remains a challenge for the cost-effective design of compact fiber lasers. Here, we exploit the concept of coherent perfect absorption to overcome this challenge, whereby two appropriately designed fiber Bragg gratings define a short-length erbium-doped-fiber cavity that enforces complete absorption of an incident field on resonance -- \textit{independently} of the doped-fiber intrinsic absorption. This approach applies to any spectral window and guarantees the efficient utilization of the fiber dopants along its length, thus suggesting the possibility of next-generation efficient single-longitudinal-mode fiber lasers for applications in optical communication, sensing, and metrology.\end{abstract}



\section{Introduction}
Coherent perfect absorption (CPA) is an interferometric phenomenon whereby a partially absorbing material is made to perfectly absorb incoming light by properly engineering the photonic environment around the absorptive medium. The effect can be explained through the combined effect of interference and critical coupling, which results in a maximal interaction between the absorber and the electromagnetic field. The initial concept of CPA was proposed and demonstrated one-dimensional (1D) structures in a two-sided-incidence configuration \cite{Chong10PRL,Wan11Science}. In this scheme, the absorbing layer is placed inside the standing wave formed from two symmetric counterpropagating beams with a specific relative phase. By tuning this relative phase, the absorption is modulated from near zero to near unity at discrete resonant wavelengths. Although this scheme is useful for optical modulation and switching \cite{Wan11Science, Papaioannou16APLPhotonics, Rao14OL, Roger15NatCommun, Rothenberg16OL, Fannin16IEEEPJ, Zhao16PRL, Papaioannou16LSA, Fang14APL}, it may not be a convenient one for other applications such as optical pumping and solar cells. It has been shown that by properly devising an asymmetric cavity around the absorbing medium, the CPA effect is achievable in a one-sided-incidence configuration \cite{Villinger15OL, Pye17OL, Li14SciRep}. The concept has been extensively applied to various materials including silicon \cite{Wan11Science, Rothenberg16OL, Fannin16IEEEPJ, Pye17OL}, patterned conductors \cite{Papaioannou16APLPhotonics, Roger15NatCommun, Papaioannou16LSA, Fang14APL, Yoon12PRL}, and 2D materials \cite{Rao14OL, Jang14PRB, Zhu16LSA, Liu14APL, Huang16ACSNano, Pirruccio13ACSNano, Thareja15NanoLett, Yao14NanoLett, Furchi12NanoLett, Kakenov16ACSPhotonics, Woo14APL}, and demonstrated in different platforms such as planar geometries \cite{Pye17OL, Jang14PRB, Furchi12NanoLett, Kakenov16ACSPhotonics, Woo14APL, Kats12APL}, on-chip systems \cite{Rothenberg16OL, Zhao16PRL, Wong16NatPhoton}, photonic crystals \cite{Zhu16LSA,Gan12NanoLett}, ultrathin metasurfaces \cite{Papaioannou16APLPhotonics, Roger15NatCommun, Papaioannou16LSA, Fang14APL, Yao14NanoLett} plasmonics \cite{Papaioannou16APLPhotonics, Yoon12PRL, Jang14PRB, Jung15OE}, and nonlinear optics \cite{Zheng13SciRep}.

Despite extensive theoretical studies of CPA and its realization in various photonic platforms, no investigations of fiber-based systems have yet been reported. Active fibers are building blocks for fiber amplifiers and fiber lasers which are widely utilized for their salutary features such as obviating the need for optical alignment, minimal mode distortion, and high efficiency \cite{Richardson10JOSAB}. Recently, compact short-length fiber lasers with single-longitudinal-mode operation have been developed for applications in coherent communication \cite{Beppu15OE}, optical clocks \cite{Foreman07RSI}, video transmission \cite{Lee98OL}, spectroscopy \cite{Kim99AO}, and sensing \cite{Perez-Herrera12JLT}. Constructing an efficient short-length fiber laser using germanosilicate fibers remains a challenge due to fundamental limitations on high pumping power stemming from difficulties associated with high ion concentrations \cite{Agrawal89NFO,Delevaque93PTL,Wagener93OL}.



This challenge prompts the following question: is it possible to enforce total absorption of the pump energy in a moderately-doped few-centimeter-long fiber \textit{independently} of the fiber's \textit{intrinsic} absorption? We propose a solution to this challenge by exploiting the notion of CPA realized here in a short-length Er-doped single-mode fiber (EDF). Despite the low intrinsic absorption of this EDF ($\approx\!7$~dB/m), we demonstrate that a 10-cm-long segment can \textit{fully} absorb an incident optical field when appropriate fiber Bragg-gratings (FBGs) are provided at the two ends of the EDF. Indeed, CPA severs the link between the intrinsic absorption in the EDF segment and the net absorption in the EDF cavity, guaranteeing full absorption on resonance. We also highlight a unique feature of the CPA configuration: incident light is fully absorbed while simultaneously optimizing the utilization of the doped active ions by maintaining a uniform axial absorption profile. This is particularly the case when compared to other traditional pumping configurations, including a bare fiber, dual-side pumping, and a fiber provided with a back-reflector. 

\begin{figure}[t!]
\begin{center}
\includegraphics[width=13.335cm]{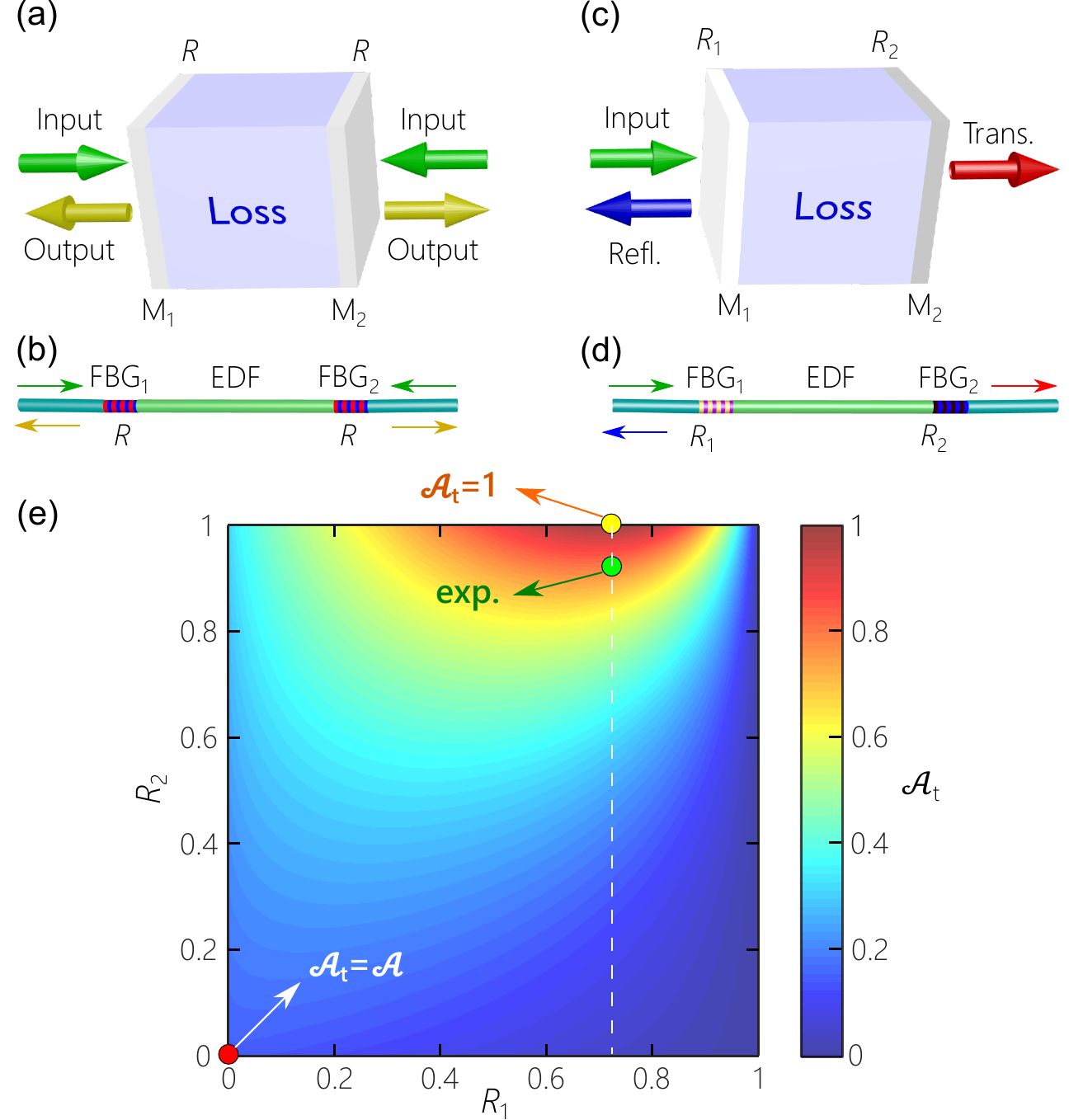}
\end{center}
\caption{\label{fig:Concept}(a) Two-sided-incidence configuration in a symmetric ($R_{1}\!=\!R_{2}\!=\!R$) planar cavity containing an absorbing layer. (b) A fiber-based cavity corresponding to (a). FBG: fiber-Bragg grating; EDF: Er-doped fiber. (c) One-sided-incidence configuration in an \textit{a}symmetric planar cavity ($R_{1}\!\neq\!R_{2}$). (d) A fiber-based cavity corresponding to (c). (e) Calculated normalized net cavity absorption $\mathcal{A}_{\mathrm{t}}$ as a function of $R_{1}$ and $R_{2}$ when $\mathcal{A}\!=\!0.15$. The vertical dashed white line corresponds to $R_{1}\!=\!(1-\mathcal{A})^{2}\!=\!0.72$. The yellow dot corresponds to the CPA condition $R_{1}\!=\!0.72$, $R_{2}\!=\!1$, and $\mathcal{A}_{\mathrm{t}}\!=\!1$; the green dot to the experimental value of $R_{2}\!\approx\!0.93$ that limits $\mathcal{A}_{\mathrm{t}}$ to $0.75$; and the red dot to the condition $\mathcal{A}_{\mathrm{t}}\!=\!\mathcal{A}$ when $R_{1}\!=\!R_{2}\!=\!0$.}
\end{figure}

\section{CPA cavity design}
We model the single-transverse-mode fiber cavity as a planar system, which allows us to employ the analysis developed in Ref.~\cite{Villinger15OL}. Consider a cavity where two mirrors M$_1$ and M$_2$ (reflectivities $R_{1}$ and $R_{2}$, respectively) sandwich a lossy layer having an \textit{intrinsic} normalized single-pass power absorption $0\!<\!\mathcal{A}\!<\!1$ [Fig.~\ref{fig:Concept}(a)]. A fiber-based realization consists of an absorbing fiber, such as an EDF, between two FBGs [Fig.~\ref{fig:Concept}(b)]. In the CPA model described in Ref.~\cite{Chong10PRL}, counter-propagating fields are incident on the two cavity ports. The CPA limit in this two-sided-incidence arrangement corresponds to no outgoing fields (both incident fields are fully absorbed), resulting in a normalized \textit{net} cavity absorption of $\mathcal{A}_{\mathrm{t}}\!=\!1$. In a symmetric cavity $R_{1}\!=\!R_{2}\!=\!R$, the following conditions are necessary for CPA: (1) the incident fields have equal amplitudes; (2) selecting $R\!=\!1-\mathcal{A}$ such that the cavity reflection and transmission coefficients for an incident field have equal amplitudes; and (3) the relative phase between the incident fields is either 0 or $\pi$ \cite{Villinger15OL}. The latter requirement is particularly challenging in practice. If one field is incident on this symmetric cavity, which eliminates the need for maintaining phase stability, it can be shown that an optimal $\mathcal{A}_{\mathrm{t}}\!=\!\frac{(2-\mathcal{A})^{2}}{8(1-\mathcal{A})}$ is achieved by selecting $R\!=\!\frac{(2-3\mathcal{A})}{(1-\mathcal{A})(2+\mathcal{A})}$ \cite{Villinger15OL}. These results apply for values of $\mathcal{A}$ in the range $0\!<\!\mathcal{A}\!<\!2/3$, beyond which the cavity yields no improvement in $\mathcal{A}_{\mathrm{t}}$ over $\mathcal{A}$. This indicates that a minimum of $\mathcal{A}_{\mathrm{t}}\!=\!0.5$ is always achievable when a single field is incident on a symmetric cavity \textit{independently} of $\mathcal{A}$ (even $\mathcal{A}\!\ll\!1$) -- as long as an appropriate value for $R$ is used.

Here we eliminate the phase-stability requirement in the two-sided-incidence configuration [Fig.~\ref{fig:Concept}(a,b)] and \textit{still} achieve CPA by `folding' the system back on itself, whereupon a strong reflector is used for M$_2$ and a single beam is incident on M$_1$ [Fig.~\ref{fig:Concept}(c)]. The corresponding asymmetric fiber-based realization requires FBGs of different reflectivities [Fig.\ref{fig:Concept}(d)]. To show that this one-port system achieves CPA, we obtain the transmission and reflection coefficients:
\begin{eqnarray}
{T}_\textrm{t}&=&\frac{(1-R_1)(1-R_2)\varGamma}{(1-\varGamma \sqrt{R_1 R_2})^2+4\varGamma\sqrt{R_1 R_2}\textrm{sin}^2\phi},\label{eq:T}\\
{R}_\textrm{t}&=&\frac{R_1+R_2\varGamma^2-2\sqrt{R_1 R_2}\textrm{cos}2\phi}{(1-\varGamma \sqrt{R_1 R_2})^2+4\varGamma\sqrt{R_1 R_2}\textrm{sin}^2\phi}\label{eq:R},
\end{eqnarray}
where $\varGamma\!=\!1-\mathcal{A}\!=\!e^{-\alpha L}$ is the single-pass power attenuation, $\alpha$ is the attenuation coefficient, $\phi\!=\!nk_{0}L+(\gamma_{1}+\gamma_{2})/2$ is the single-pass phase, $n$ is the refractive index, $L$ is the cavity length, and $\gamma_{1}$ and $\gamma_{2}$ are the reflection phases of M$_1$ and M$_2$, respectively. Therefore, the net cavity absorption $\mathcal{A}_\textrm{t}\!=\!1-{T}_\textrm{t}-{R}_\textrm{t}$ is
\begin{equation}\label{eq:TotalAbsorptionGeneral}
\mathcal{A}_\textrm{t} = \frac{(1-R_1)(1-\varGamma)(1+\varGamma R_2)}{(1-\varGamma \sqrt{R_1 R_2})^2+4\varGamma\sqrt{R_1 R_2}\textrm{sin}^2\phi}.
\end{equation}
It is straightforward to prove that perfect absorption -- for a given $\varGamma$ -- is predicated on satisfying the following conditions: (1) the back mirror M$_2$ is a perfect reflector $R_{2}\!=\!1$; (2) the front mirror M$_1$ reflectivity is $R_{1}\!=\!\varGamma^2$; and (3) the incident-field wavelength coincides with a cavity resonance, $\phi\!=\!m\pi$ for integer $m$. Substituting $R_{1}\!=\!\varGamma^{2}$ and $R_{2}\!=\!1$ in Eq.~\ref{eq:TotalAbsorptionGeneral} yields
\begin{equation}
\mathcal{A}_{\textrm{t}}=\left(1+\frac{2\varGamma}{1-\varGamma^2}\sin{\phi}\right)^{-2},
\end{equation}
whereupon $\mathcal{A}_{\textrm{t}}\!=\!1$ on resonance ($\sin{\phi}\!=\!0$) \textit{independently} of $\varGamma$.

To illustrate this enhancement in cavity absorption, we plot $\mathcal{A}_{\mathrm{t}}$ in Fig.~\ref{fig:Concept}(e) as a function of $R_1$ and $R_{2}$ at a single-pass absorption of $\mathcal{A}\!=\!0.15$ (the value used in our experiment). When $R_{1}\!=\!R_{2}\!=\!0$, we have $\mathcal{A}_{\mathrm{t}}\!=\!\mathcal{A}$ [the red dot in Fig.~\ref{fig:Concept}(e)]. The CPA case $\mathcal{A}_{\mathrm{t}}\!=\!1$ is achieved when $R_{2}\!=\!1$ and $R_{1}\!=\!\varGamma^2\!=\!0.72$ [the yellow dot in Fig.~\ref{fig:Concept}(e)]. Note that $\mathcal{A}_{\mathrm{t}}$ drops quickly when $R_{2}$ departs from unity, but is less sensitive to deviations in $R_{1}$.

\begin{figure}[t!]
\begin{center}
\includegraphics[width=13.335cm]{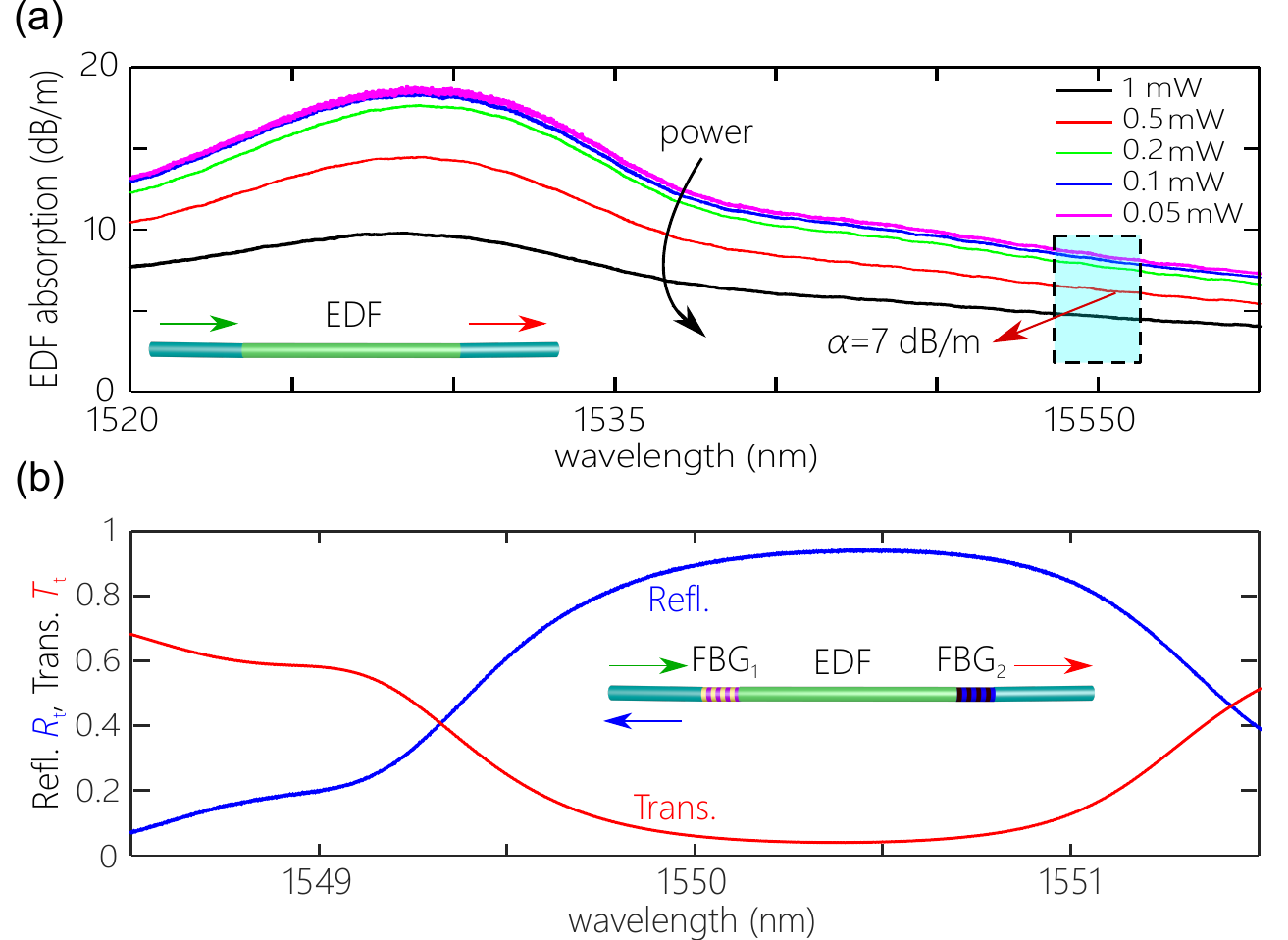}
\end{center}
\caption{(a) Measured absorption in the EDF (M-15; Fibercore, Ltd.) at different power levels. To achieve CPA, the power level is chosen such that the single-pass EDF absorption $\mathcal{A}$ corresponds to $\sqrt{1-R_{1}}$. (b) Measured transmission and reflection of the EDF cavity with low spectral-resolution over a $\approx\!3$-nm bandwidth corresponding to the highlighted range in (a). Measurements are carried out using a wide-band ASE source.
\label{fig:SampleData}}
\end{figure}

As a proof-of-principle experimental demonstration of CPA in an EDF, we construct a short-length C-band fiber cavity. We make use of an EDF (M-15; Fibercore, Ltd.) that provides an intrinsic absorption of $\alpha\!\approx\!7$~dB/m at a wavelength 1550~nm, and an effective mode-field-diameter of $\SI{6.3}{\micro m}$ ($\pm\SI{0.6}{\micro m}$). A weakly absorbing $L\!=\!10$-cm-long fiber segment ($\approx\!60,000\times$ the incident wavelength) with $\mathcal{A}\!\approx\!0.15$ requires a front FBG having a reflectivity $R_{1}\!\approx\!0.72$. Such a cavity has a free spectral range (FSR) of $\approx\!8$~pm (FWHM) at the operating wavelength. Note that the EDF absorption changes with the incident power \cite{Myslinski97JLT}. At low power, only a fraction of the active ions are excited and the the absorption is determined by the ion concentration. At high power, a significant fraction of the ions get excited, leading to absorption saturation -- an effect that is more prominent at low concentrations \cite{Kiryanov13JQE}. As a result, the effective fiber absorption decreases with increasing incident power, as confirmed by our measurements in Fig.~\ref{fig:SampleData}(a). The EDF absorption of $\alpha\!\approx\!7$~dB/m is measured at a power of $\SI{500}{\micro W}$. We therefore hold this power level fixed throughout the experiment. Finally, note that the C-band was selected for convenience, but the principle can be readily extended to other bands and is applicable to any rare-earth ions or other dopants.

The cavity mirrors are custom-made FBGs (O-Eland, Inc.) with lengths $<\!1$~mm ($\ll\!L$) written directly in the EDF to maintain a minimum distance between the FBGs and eliminate any splicing losses. Both FBGs have a central wavelength of $\approx\!1550.3$~nm and a $3$-dB bandwidth of $\approx\!2$~nm. These spectral features are confirmed from low spectral-resolution measurements of $R_{\mathrm{t}}$ and $T_{\mathrm{t}}$ for the cavity as shown in Fig.~\ref{fig:SampleData}(b). The target values in our experiment are $\mathcal{A}\!\approx\!0.15$ are $R_{1}\!\approx\!0.72$ and $R_{2}\!=\!1$ [yellow dot in Fig.~\ref{fig:Concept}(e)]. Difficulties in writing the FBG directly in the EDF have limited $R_{2}$ to 0.93, which in turn limits the maximum achievable value of $\mathcal{A}_{\mathrm{t}}$ to $\approx\!0.75$ [green dot in Fig.~\ref{fig:Concept}(e)]. Note that absorption saturation in the EDF enables us to modify $\mathcal{A}$ by tuning the power to optimally match the value of $R_{1}$ of the front FBG.

\section{Measurements and results}
\begin{figure}[p]
\begin{center}
\includegraphics[width=12.cm]{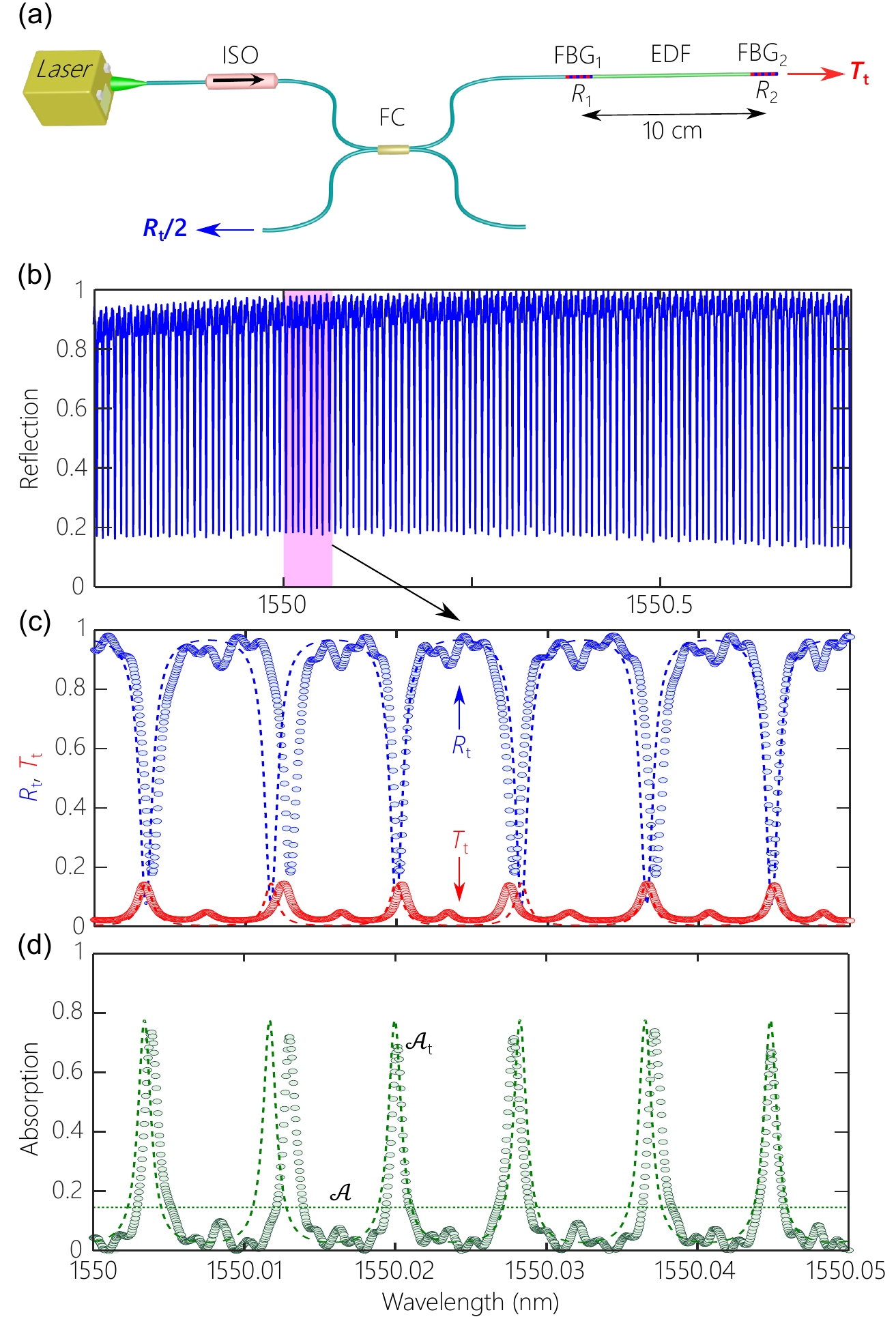}
\end{center}
\caption{Experimental demonstration of CPA in a short-length EDF. (a) Schematic of the setup to measure the spectral transmission $T_{\mathrm{t}}$ and reflection $R_{\mathrm{t}}$ from the EDF-cavity. ISO: isolator; FC: 50:50 fiber coupler. (b) Measured $R_{\mathrm{t}}$ across a 1-nm bandwidth ($>\!100$~FSRs). (c) Measured $R_{\mathrm{t}}$ and $T_{\mathrm{t}}$ and (d) $\mathcal{A}_{\mathrm{t}}$ in a 50-pm bandwidth, corresponding to the highlighted span in (b). The data points in (b,c) are shown as circles and the dashed lines are theoretical predictions. The dotted horizontal line is the measured value of $\mathcal{A}$ in the EDF. The best fit between experiment and theory is achieved when setting $\mathcal{A}\!\approx\!0.19$.\label{fig:SetupAndData}}
\end{figure}

We carry out the CPA measurements in the EDF using the optical arrangement illustrated in Fig.~\ref{fig:SetupAndData}(a). The setup enables measuring the cavity spectral reflection $R_{\mathrm{t}}$ and transmission $T_{\mathrm{t}}$ (both normalized with respect to the source spectrum), from which we obtain the absorption $\mathcal{A}_{\mathrm{t}}$. A tunable laser diode (Agilent 81640A; linewidth 100~kHz) sweeps across a 3-nm-wide spectrum (from 1548.5 to 1551.5~nm) at a sweeping speed of 0.5~nm/s in wavelength steps of 0.1~pm. The relfection and transmission ports are connected to fiber-coupled detectors (Newport 818-IR-L) that sample the data at a 10-kHz rate ($\approx\!20$ data points in the FWHM of each resonance). The measured spectral $R_{\mathrm{t}}$ and $T_{\mathrm{t}}$ are plotted in Fig.~\ref{fig:SetupAndData}(b,c). In the ideal CPA scenario, we have $R_{\mathrm{t}}\!=\!1$ and $T_{\mathrm{t}}\!=\!0$, so that $\mathcal{A}_{\mathrm{t}}\!=\!1$. The data however shows that at the cavity resonances there are remnants of reflection and transmission as a result of back-reflector being less than unity ($R_{2}\!\approx\!93\%$).
The theoretical plots utilize Eqs.~\ref{eq:T},\ref{eq:R} after substituting the experimental values of the parameters. Using a shorter fiber or lower ion concentration reduces $\mathcal{A}$, which requires a larger value of $R_{1}$, but $\mathcal{A}_{\mathrm{t}}\!=\!1$ is always guaranteed as long as the correct value of $R_{1}$ is implemented.

\begin{figure}[p]
\begin{center}
\includegraphics[width=12.cm]{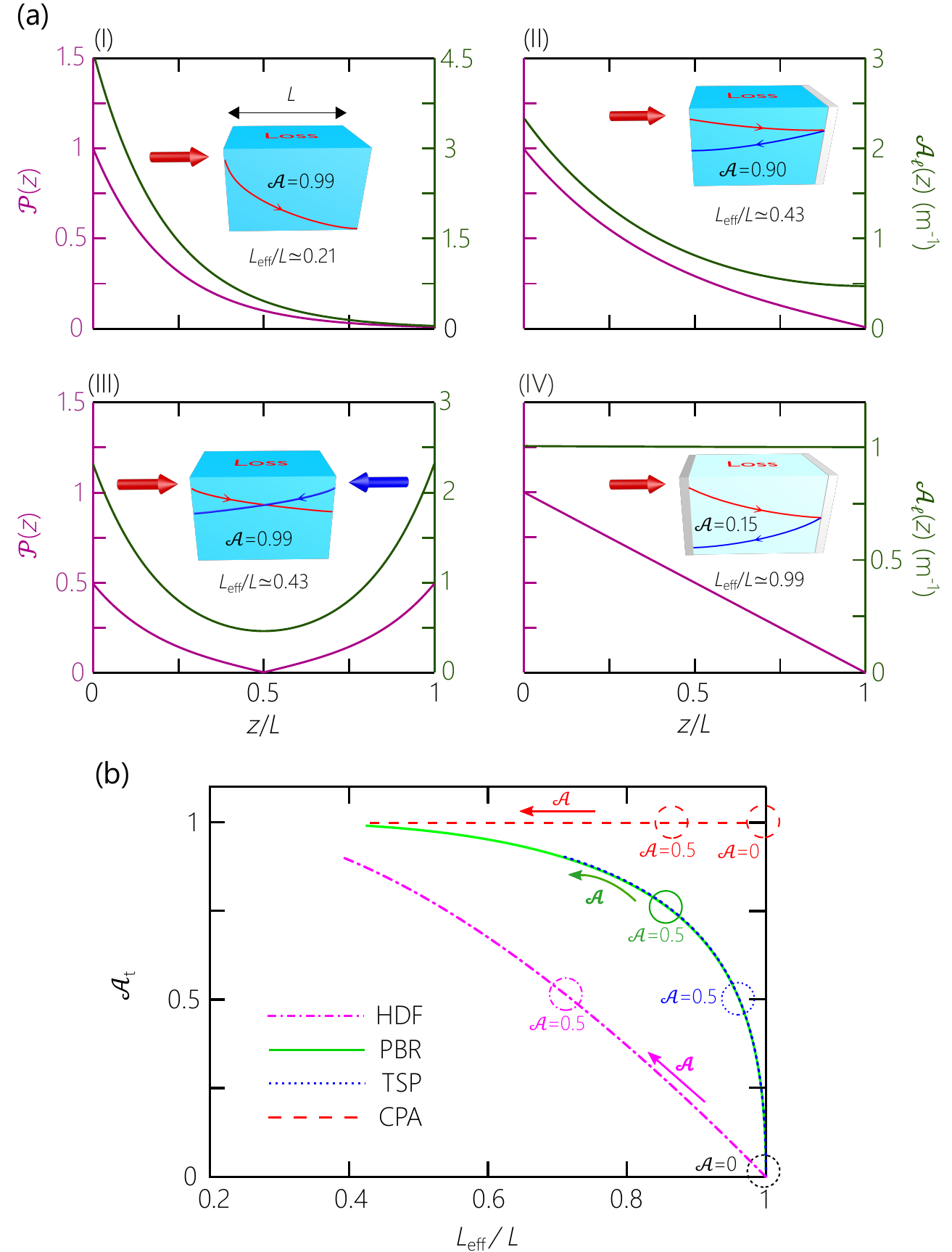}
\end{center}
\caption{(a) Plots of the axial component of Poynting's vector $\mathcal{P}(z)$ normalized with respect to the total incident on the cavity (magenta curves, axis to the left), and the local absorption per unit length $\mathcal{A}_{\ell}(z)=-\partial\mathcal{P}(z)/\partial z$ (green curves, axis to the right) -- along the normalized axial coordinate $z/L$. Four configurations are depicted (shown schematically as insets): (I) single pass in a heavily doped fiber (HDF); (II) fiber provided with a perfect back-reflector (PBR); (III) two-sided pumping (TSP) with equal-amplitude counterpropagating fields; and (IV) CPA. The total fiber absorption in designed in each configuration to be $\mathcal{A}_\mathrm{tot}>99\%$. Under this condition, the single-pass absorption $\mathcal{A}$ and the effective fiber length $L_\mathrm{eff}\!/\!L$ (which quantifies the utilization of the fiber dopants) are listed in each panel. (b) Parametric plots of $\mathcal{A}_{\mathrm{t}}$ versus $L_{\mathrm{eff}}\!/\!L$ as $\mathcal{A}$ is varied for the configurations shown in (a). The arrows show the direction of increasing $\mathcal{A}$ from 0 to 0.90.
\label{fig:Comparison}}
\end{figure}

In addition to increasing $\mathcal{A}_{\mathrm{t}}$ on resonance, potentially to $100\%$ independently of $\mathcal{A}$, the CPA configuration has a useful advantage that has not been previously noted. Namely, the \textit{axial}-dependence of absorption along the EDF is considerably more uniformly distributed in stark contrast to the axial-absorption profile in a heavily doped EDF. Therefore, CPA allows for the most efficient utilization of the fiber dopants. The power absorption is quantified as the power drop per unit length along the cavity, $\mathcal{A}_{\ell}(z)\!=\!-\,\partial\mathcal{P}(z)/\partial z$, where $\mathcal{P}(z)$ is Poynting's vector magnitude normalized at the axial position $z\!=\!0$. We quantify the utilization of the EDF as an absorber by defining an effective length $L_{\mathrm{eff}}$,
\begin{equation}
L_{\mathrm{eff}}=\int_{0}^{L}\,\mathcal{A}_\mathrm{norm}(z)\;\mathrm{d}z,
\end{equation}
where $\mathcal{A}_\mathrm{norm}(z)$ is the local absorption per unit length normalized to the absorption at $z\!=\!0$. When $L_{\mathrm{eff}}\!=\!L$, absorption takes place uniformly along the fiber, and the EDF is efficiently utilized, while $L_{\mathrm{eff}}\!<\!L$ indicates a non-uniform absorption distribution that concentrates the absorption in sub-sections of the EDF. Here $L_{\mathrm{eff}}$ quantifies only the axial uniformity of the absorption distribution, especially for moderate values of $\mathcal{A}$, and more complex measures are needed to account simultaneously for absorption saturation at high power levels.

To highlight this feature quantitatively, we compare four pertinent configurations -- each comprising a fiber with single-pass absorption $\mathcal{A}$ [Fig.~\ref{fig:Comparison}(a)]. (I) A single pass through a heavily doped EDF corresponding to $R_{1}\!=\!R_{2}\!=\!0$ and $\mathcal{A}_{\mathrm{t}}\!=\!\mathcal{A}$, such that $L_{\mathrm{eff}}/L\!=\!(\varGamma-1)/\ln\varGamma$, where $\varGamma\!=\!1-\mathcal{A}$ is the single-pass attenuation. It is possible to reach a high net absorption by increasing $\mathcal{A}$, at the expense of inefficient utilization of the EDF ($L_{\mathrm{eff}}\!<\!L$). The field is absorbed mostly in the first section of the fiber and the doped ions at the other end of the fiber are less utilized. (II) An EDF with a back-reflector corresponding to $R_{1}\!=\!0$, $R_{2}\!=\!1$, $\mathcal{A}_{\mathrm{t}}\!=1-\varGamma^2\!>\!\mathcal{A}$, in which case $L_{\mathrm{eff}}/L\!=\!\frac{-1}{\ln\varGamma}\frac{1+\varGamma^2}{1-\varGamma^2}$; (III) an EDF with two counter-propagating equal-amplitude fields, where $R_{1}\!=\!R_{2}\!=\!0$ and $\mathcal{A}_{\mathrm{t}}\!=\!\mathcal{A}$ and $L_{\mathrm{eff}}/L\!=\!\frac{-2}{\ln\varGamma}\frac{1+\varGamma}{1-\varGamma}$. Symmetrization of absorption along the fiber length can be improved in configurations (II) and (III). This symmetrization is best achieved at low $\mathcal{A}$ (and thus low $\mathcal{A}_{\mathrm{t}}$), whereupon $L_{\mathrm{eff}}\!\rightarrow\!L$. Increasing $\mathcal{A}$ increases $\mathcal{A}_{\mathrm{t}}$ but reduces $L_{\mathrm{eff}}$. 
(IV) The CPA configuration, where $R_{1}\!=\!(1-\mathcal{A})^{2}$, $R_{2}\!=\!1$, $\mathcal{A}_{\mathrm{t}}\!=\!1$, and $L_{\mathrm{eff}}$ is the same as configuration (II) with a back-reflector. The four configurations (I) through (IV) in Fig.~\ref{fig:Comparison}(a) are designed such that the fiber system has a total absorption of $\mathcal{A}_{\mathrm{t}}\!>\!99\%$. We note two advantages of the CPA configuration: a small single-pass absorption $\mathcal{A}$ is required, and the axial absorption profile is uniform with respect to the other three more familiar configurations.

The results for the four absorption configurations are summarized in Fig.~\ref{fig:Comparison}(b) where we plot parametric curves of $\mathcal{A}_{\mathrm{t}}$ against $L_{\mathrm{eff}}\!/\!L$ while varying $\mathcal{A}$. These curves bring out the advantages of the CPA configuration clearly. In particular, at low single-pass absorption $\mathcal{A}\!\rightarrow\!0$, CPA combines perfect absorption $\mathcal{A}_{\mathrm{t}}\!=\!1$ with a large effective length $L_{\mathrm{eff}}\!\rightarrow\!L$ (and thus highly efficient utilization of the fiber dopants).

Finally, we comment on the potential utilization of CPA in constructing a fiber laser. A CPA cavity would be designed for efficient absorption at the pump wavelength, while a secondary cavity would be introduced for the emission wavelength. Such a scheme can be implemented with an additional pair of FBGs written in the doped fiber section \cite{Becker99AP}. Open questions remain for future research. Critically, does the enhancement in $\mathcal{A}_{\mathrm{t}}$ produce a corresponding enhancement in the fiber gain? Furthermore, instead of the core-pumping scheme explored here, can the CPA approach be extended to cladding-pumping arrangements where the multimode operation can help with resonance multiplexing? 

\section{Conclusion}
In conclusion, we have presented -- to the best of our knowledge -- the first demonstration of CPA in a short-length, low-absorption EDF, which may have important implications for compact fiber laser systems. Such a general scheme may help increase the lasing efficiency by improving pump utilization. Crucially, this strategy is not limited to a specific ion or wavelength range, and is readily implemented by providing a pair of suitably designed FBGs to a weakly absorbing fiber section. Furthermore, the CPA configuration guarantees optimally utilizing the EDF by maintaining uniform local absorption along the fiber. Single-longitudinal-mode lasing action may be possible in such short-length fibers if the FBG bandwidths are sufficiently narrow. Such narrow-linewidth compact systems can help address stringent requirements for high temporal coherence and insensitivity to environment. Note that the design in its current form requires further development to be included in a fiber laser. Most notably, the linewidth of the resonant absorption in the current embodiment is considerably narrower than the bandwidth of commercially available diode lasers. Further research is thus needed to identify an approach for broadening the cavity resonances. One avenue that is particularly attractive is to multiplex resonances in a multi-mode fiber arrangement.

\section*{Funding}
Air Force Office of Scientific Research (AFOSR) under MURI award
FA9550-14-1-0037.

\section*{Acknowledgments} The authors are grateful to Amy Van Newkirk for generous assistance and useful discussions, and to Guifang Li and He Wen for loan of equipment.

\end{document}